\documentclass[11pt]{article}
\usepackage{amssymb,amsmath,amsfonts}
\usepackage{graphicx}
\usepackage{graphics}
\usepackage{eepic,epsfig}

\begin{document}

\title{Cylindrical boundary induced current in the cosmic string spacetime}

\author{E. R. Bezerra de Mello$^1$\thanks{emello@fisica.ufpb.br} \ , 
H. F. Santana Mota$^1$\thanks{hmota@fisica.ufpb.br} and W. Oliveira dos Santos$^1$\thanks{E-mail: wagner.physics@gmail.com}
	\\
	\textit{$^{1}$Departamento de F\'{\i}sica, Universidade Federal da Para\'{\i}ba,}\\
	\textit{58.059-970, Caixa Postal 5.008, Jo\~{a}o Pessoa, PB, Brazil}}

\maketitle

\begin{abstract}
In this paper we investigate the vacuum current associated with a charged bosonic field operator, induced by a cylindrical boundary in the idealized cosmic string spacetime. In this setup we assume that the cylindrical boundary is coaxial with the string, that by its turn carry a magnetic flux along its core. In order to develop this analysis, we calculate the positive frequency  Wightman functions for both regions, inside and outside the boundary. Moreover, we assume that the bosonic field obeys the Robin boundary condition on the cylindrical shell. Using this approach, the analytical expressions for the vacuum bosonic currents are presented in the form of the sum of boundary-free and boundary-induced parts. Because the boundary-free contribution is very well established in literature, our focus here is in the boundary-dependent part. As we will see, our general  results are presented in a cylindrically symmetric static structure. Some asymptotic behaviors for the boundary-induced vacuum currents are investigated in various limiting cases. In order to provide a better understanding of these currents, we provide some graphs exhibiting their behavior as function of the distance to the string's core, and on the intensity of the magnetic flux running along it. These plots also present how the parameter associated with the planar angle deficit interfere in the intensity of the corresponding current.  
\end{abstract}

\bigskip


\section{Introduction}
Cosmic strings are linear topological objects that may be formed as consequence of gauge symmetry breaking in early stage of the Universe evolution \cite{Kibble,V-S}. The gravitational field produced by a cosmic string can be approximated by a planar angle deficit in the two-dimensional sub-space orthogonal to the string.  Although in recent observational data on the cosmic microwave background have discarded cosmic strings as the primary source for primordial density perturbation, these object are still candidate for the generation of interesting physical effects such as gamma ray bursts \cite{Berezinski}, gravitational waves \cite{Damour} and high energy cosmic rays \cite{Bhattacharjee}. Recently, cosmic strings have attracted renewed interest partly because a variant of their formation mechanism is proposed in the framework of brane inflation \cite{Sarangi}-\cite{Dvali}.

Under quantum field theory viewpoint, the conical topology associated with the cosmic string leads to non-zero vacuum expectation values for physical observables. In this sense the vacuum expectation values (VEVs)  of physical observables, like the energy-momentum tensor, $\langle T_{\mu\nu}\rangle$, are calculated considering scalar and fermionic quantum fields. (See  \cite{PhysRevD.35.536, escidoc:153364, GL, DS, PhysRevD.46.1616,PhysRevD.35.3779, LB, BK}). Furthermore, taking into account the presence of a magnetic flux running along string's core, additional contributions to the VEVs  associated with charged fields \cite{PhysRevD.36.3742, guim1994, SBM, SBM2, SBM3, Spinelly200477, SBM4} takes place. Another important observable to be considered in this system is the induced vacuum current densities, $\langle j^\mu\rangle$. This phenomenon has been investigated for massless \cite{LS}, massive scalar \cite{SNDV} and fermionic \cite{ERBM} fields, respectively. In these papers, the authors have shown that induced vacuum current densities along the azimuthal direction arise if the ratio of the magnetic flux by the quantum one has a nonzero fractional part. Moreover, the induced bosonic current in higher-dimensional compactified cosmic string spacetime was calculated in \cite{Braganca2015}.

The presence of boundaries also produce modification on the VEV of physical observables due to a changing on the physical properties of vacuum state. This is the well-known Casimir effects. The analysis of Casimir effects in the idealized cosmic string space-time have been developed for scalar, fermionic and vector fields in \cite{Mello,Mello1,Aram1} obeying specific boundary conditions on cylindrical surface. Specifically in \cite{Mello} it was analyzed bosonic vacuum polarization induced by a cylindrical shell in a high dimensional cosmic string spacetime. By imposing Robin boundary condition on the field in the cylindrical shell, the VEVs of the energy-momentum tensor and field squared were analyzed for the both regions: inside and outside the shell.  Moreover, the analysis of Casimir effects induced by just one flat boundary orthogonal to the string have been developed for scalar and fermionic fields in \cite{MelloCQG2001} and \cite{MelloCQG2013}, respectively, and for the case of two flat boundaries in \cite{MelloPRD2018}. Considering the presence of charged bosonic field in this background, in \cite{Braganca:2021} it was calculated the VEV of the energy-tensor, $\langle T_{\mu\nu}\rangle$, and induced current, $\langle j^\mu\rangle$.

Vacuum currents induced by topological defects and boundaries are of interest in both high energy  and condensed matter physics and similar geometries appear in models
of superconducting strings \cite{Mankiewicz_a,Mankiewicz_b}, flux tubes, graphene \cite{Sitenko,Liwei}, and other defect-based systems \cite{Muniz,Sitenko:2019lug}. More specifically, in cosmology for instance, cosmic strings may interact with background gauge fields and surfaces, and the resulting vacuum polarization effects can influence field configurations in early-universe models \cite{V-S}. In condensed matter systems, analogues of conical geometry appear in materials such as graphene \cite{Sitenko,Liwei} or defect-rich topological insulators \cite{Muniz,Sitenko:2019lug}, where edge-induced currents are observable. Motivated by these settings, we analyze how a cylindrical boundary modifies the induced vacuum current in a cosmic string background threaded by a magnetic flux.

As we have mentioned before, the analysis of the influence of a cylindrical boundary on the VEV of the energy-momentum tensor associated with scalar quantum field in higher-dimensional cosmic string space-time was developed in \cite{Mello}. In this way, the latter completes the study  on the calculation of VEV  of the energy-momentum tensor associated with scalar field. In addition, the analysis of azimuthal induced bosonic current in higher-dimensional compactified cosmic string space-time was developed in \cite{Braganca2015}. However, an important discussion has been missed so far, that is the analysis of the influence of cylindrical boundary on the induced bosonic current. The main objective of this paper is to fill out this gap. Having obtained the corresponding analytical expressions, we intend to analyze their most important features, such as their dependences with the distance. 

This paper is organized as follows: In Section \ref{sec2} we present the background geometry associated with a high dimensional cosmic string  spacetime, and the explicit expression for the four-vector potential associated with the magnetic flux. Considering the Klein-Gordon equation and the Robin boundary condition imposed to the charged field at the cylindrical shell, we calculate the complete set of normalized wave-function. By using the mode summation formula, we obtain the positive frequency Wightman functions for both regions of the space. As we will see the corresponding Wightman functions are expressed in terms of the boundary-free contributions plus a boundary-induced contribution. Because our main objective here is to calculate the boundary-induced current, our focus is in the second contribution.
In Section \ref{sec3} we investigate the vacuum bosonic currents induced by the boundary for both regions. There we will see that only azimuthal currents are induced. Also we provide some asymptotic behaviors for these currents. In Section \ref{conc}  we summarize the most relevant results obtained. In  this paper we will use the units $\hbar =G=c=1$.

\section{Wightman function}
\label{sec2}
In this section we will investigate the quantum behavior of a charged field in a high-dimensional  cosmic string spacetime, considering the presence of a magnetic flux along its core, and the presence of a cylindrical boundary coaxial with the string. On its surface we impose that the field obeys the Robin boundary condition. Ou main objective is to calculate the positive frequency Wightman functions considering both regions of the space. 

\subsection{Bulk geometry and bosonic modes} 
In this subsection we present the $(D+1)$-dimensional background spacetime with a conical-type singularity described
by the line-element below:
\begin{equation}
	ds^{2}=g_{\mu\nu}dx^{\mu}dx^{\nu}=dt^{2}-dr^{2}-r^{2}d\varphi
	^{2}-\sum_{i=3}^{D}dx_{i}{}^{2},  \label{ds21}
\end{equation}
with the cylindrical coordinates $r\geqslant 0$, $0\leqslant \varphi\leqslant{2\pi}/{q}$, and  $-\infty <x_{i}<+\infty $. The presence of the cosmic string is encoded by the parameter $q\geq 1$. In the case of $D=3$, this parameter is related to the linear mass density of the string, $\mu_0$, by $q^{-1}=1-4\mu_0 $. 

The dynamics of a charged bosonic field, $\phi_\sigma (x)$, with mass $m$ in a curved spacetime and in the presence of an electromagnetic potential vector, $A_\mu$, is governed by the Klein-Gordon equation below,
\begin{equation}
	\frac{1}{\sqrt{|g|}}D_\mu\left(\sqrt{|g|}\,g^{\mu\nu}D_\nu\right)\phi_\sigma (x)+m^{2}\phi_\sigma (x)=0 \ ,
	\label{eq02}
\end{equation}
with the notation $D_{\mu}=\partial_{\mu}+ieA_{\mu}$ and $g={\rm det}(g_{\mu\nu})$. The index $\sigma$ represents the complete set of quantum numbers that characterize the wave-function. Moreover, in our analysis we assume that only the azimuthal component of the vector potential does not vanish, i.e., 
\begin{eqnarray}
A_\mu=(0, \ 0, A_\varphi, \ 0 , \ ...)	 \ ,
\end{eqnarray}
with $A_\varphi=-\frac{q\Phi}{2\pi}$,  being $\Phi$ the magnetic flux along the string. 

In addition to be solution of the Klein-Gordon equation above, we impose that the field obeys Robin boundary condition on the cylindrical surface with radius $a$, coaxial with the string:
\begin{equation}
	\left( A+B{\partial_r }\right) \phi_\sigma =0,\quad r=a.
	\label{Dirbc}
\end{equation}
Of course, all results in what follows will depend on the ratio between the coefficients in this boundary condition. However, to keep the transition to Dirichlet  ($B=0$), and Neumann ($A=0$) cases transparent, we use the more general form (\ref{Dirbc}).

In the spacetime defined by \eqref{ds21} and in the presence of the vector  potential given above, the equation \eqref{eq02} becomes
\begin{equation}
	\left[\partial_t^2-\partial_r^2-\frac{1}{r}\partial_r-\frac{1}{r^2}(\partial_{\phi}+
	ieA_{\phi})^2-\sum_{i=3}^{D}\partial_{i}^{2}+m^2\right]\phi_\sigma(x)=0 \ . 
	\label{diffe.eq}
\end{equation}

The properties of the vacuum state can be described in terms of the positive frequency Wightman function, $W(x,x')=\langle 0|\hat\phi(x)\hat\phi^\dagger(x')|0 \rangle$, where $|0 \rangle$ represents the vacuum state.  Expressing the field operator in terms of creation  and annihilation operators, the evaluation of the Wightman function is given in terms of the mode sum formula
\begin{equation}
	W(x,x')=\sum_{\sigma}\phi_{\sigma}(x)\phi_{\sigma}^{*}(x') \ .
	\label{W.function}
\end{equation}
\subsection{Wightman function in the inner region}
\label{int_reg}
Due to the cylindrical symmetry of the system under consideration, and assuming that the wave-function satisfies the Dirichlet condition at origin, $r=0$, the general solution of \eqref{diffe.eq} is given by,
\begin{eqnarray}
	\label{solu1}
	\phi_\sigma(x)=\beta_\sigma J_{q|n+\alpha|}(\gamma r)e^{-i(\omega t-iqn\phi -i{\vec{k}}\cdot {\vec{r}}_{\parallel })}  \  ,
\end{eqnarray}	
where $\mathbf{r}_{\parallel }=(x^3, \ x^4, \ ... \ x^D)$ and  $J_\mu(z)$ represents the Bessel function \cite{Abra}. In \eqref{solu1} we have
\begin{eqnarray}
	\omega&=&\sqrt{\gamma^2+{\vec{k}}^{2}+m^2} \ , \nonumber\\  
	\alpha&=&-\frac{e\Phi}{2\pi} \ .
	\label{def1}
\end{eqnarray}

According to the Robin boundary condition, the eigenvalues for the quantum number $\gamma $ are quantized. They should be compatible with the relation,
\begin{eqnarray}
	{\bar{J}}_{q|n+\alpha|}(\gamma a)=0 \  ,
	\label{Cond.1}
\end{eqnarray}
where from now on we will adopt, for any function $f(z)$, the notation, 
\begin{equation}
	\bar{f}(z)=Af(z)+\left( B/a\right) zf^{\prime }(z)  \  .  \label{fbar}
\end{equation}

Considering $\alpha$ fixed, for a given $n$ the possible values of $\gamma $ in \eqref{Cond.1} are determined by the relation
\begin{equation}
	\gamma =\lambda _{\nu_n,j}/a,\quad j=1,2,\cdots ,  \label{ganval}
\end{equation}%
where, for convenience, we will assume for while the notation $\nu_n=q|n+\alpha|$. In \eqref{ganval},  $\lambda _{\nu_n,j}$ represents the positive zeros arranged in ascending order, $\lambda _{\nu_n,j}<\lambda _{\nu_n,j+1}$, for $n=0,1,2,\ldots $. 

The coefficient $\beta_\sigma$ can be obtained by the normalization condition,
\begin{eqnarray}
	\label{Norm1}
	\int_0^a d^Dx\sqrt{|g|}g^{00}\phi^*_{\sigma'}(x)\phi_\sigma(x)=\frac1{2\omega}\delta_{\sigma',\sigma} \  , 
\end{eqnarray}
where the delta symbol on the right-hand side is understood as Dirac delta  function for the continuous quantum number ${\vec{k}}$, and Kronecker delta for the discrete ones, $n$ and $j$. From \eqref{Norm1}, and after some algebraic manipulations involving Bessel functions, we obtain
\begin{equation}
	|\beta_{\sigma}|^2=\frac{q(2\pi)^{1-D}\lambda_{\nu_n,j} T_{\nu_n}(\lambda_{\nu_n,j})}{a\sqrt{\lambda_{\nu_n,j}^2+a^2(\vec{k}^2+m^2)}}\ .
	\label{beta1}
\end{equation}
In the above expression we have for the function $T_{\nu_j}(z)$, the following result:
\begin{eqnarray}
	T_{\nu_n}(z)=z[(z^2-{\nu_n}^2)J_{\nu_n}^2(z)+z^2(J_{\nu_n}^{\prime}(z))^2]^{-1} \ .
\end{eqnarray} 

Now we are in position to calculate the Wightman function by using \eqref{W.function}. For this case we use the compact notation,
\begin{equation}
	\sum_{\sigma }=\sum_{n=-\infty}^{+\infty} \ \int d^{D-2}{k} \ \sum_{j=1}^\infty\ .  \label{Sumsig}
\end{equation}

Substituting the explicit expressions for the normalized bosonic wave-function, we obtain:
\begin{eqnarray}
W(x,x')&=&\frac{q}{(2\pi)^{D-1}a^2}\sum_{n=-\infty}^{+\infty} \ \int d^{D-2}{k} \ \sum_j\frac{\lambda_{\nu_n,j}T_{\nu_n}(\lambda_{\nu_n,j})}{\sqrt{(\lambda_{\nu_j}/a)^2+\vec{k}^2+m^2}}\nonumber\\
&\times&J_{\nu_n}(\lambda_{\nu_n,j} r/a)J_{\nu_n}(\lambda_{\nu_n,j} r'/a) e^{-i[\omega(t-t')-qn(\varphi-\varphi')-{\vec{k}}\cdot( {\vec{r}}-{\vec{r}}')_{\parallel} ]}  \ .
\end{eqnarray}

Because we do not know the explicit expressions for the eigenvalues $\lambda _{\nu_n,j}$ as functions on $n$ and $j$, the direct summation over the quantum number $j$ is not convenient. So, to obtain the Wightman function we will apply to the sum over $j$ by using a  variant of the generalized Abel-Plana summation formula \cite{Saha87}
\begin{eqnarray}
	\sum_{j=1}^{\infty }T_{\nu_n}(\lambda _{\nu_n,j})f(\lambda _{\nu_n,j}) &=&\frac{1}{2}\int_{0}^{\infty}dz\,f(z)-\frac{1}{2\pi }\int_{0}^{\infty }dz\,\frac{\bar{K}_{\nu_n}(z)}{\bar{I}_{\nu_n}(z)}  \notag \\
	&&\times \left[ e^{-\nu_n\pi i}f(iz)+e^{\nu_n\pi i}f(-iz)\right] ,  \label{sumform1AP}
\end{eqnarray}%
being $I_{\nu}(z)$ and $K_{\nu}(z)$ the modified Bessel functions \cite{Abra}. 

For our case we have,
\begin{equation}
	\label{f_function}
f(z)=\frac{zJ_{\nu_n}(z r/a)J_{\nu_n}(z r'/a)}{\sqrt{(z/a)^2+{\vec{k}}^2+m^2}}e^{-i\sqrt{(z/a)^2+{\vec{k}}^2+m^2}(t-t')} \ .
\end{equation}

Substituting \eqref{f_function} into \eqref{sumform1AP}, the Wightaman function is expressed as the sum of a boundary-free contribution, plus the boundary-induced one:
\begin{eqnarray}
	\label{W_csb}
W(x,x')=W_{cs}(x,x')+W_b(x,x') \ .
\end{eqnarray}

The boundary-free term, $W_{cs}(x,x')$, is given by the first integral in \eqref{sumform1AP}. It is given by \footnote{A more convenient expression for \eqref{W_function_cs} was obtained in \cite{Braganca:2021}.},
\begin{eqnarray}
	\label{W_function_cs}
W_{cs}(x,x')&=&\frac{q}{2(2\pi)^{D-1}}\sum_{n=-\infty}^{+\infty}e^{iqn(\varphi-\varphi')}\int d^{D-2}{k}\int d\gamma \gamma \frac{J_{q|n+\alpha}(\gamma r)J_{q|n+\alpha|}(\gamma r')}{\sqrt{\gamma^2+{\vec{k}}^{2}+m^2}}\nonumber\\
&\times& e^{-i[\omega(t-t')-qn(\varphi-\varphi')-{\vec{k}}\cdot( {\vec{r}}-{\vec{r}}')_{\parallel} ]} \  .
\end{eqnarray}
The boundary-induced contribution, $W_b(x,x')$, is  obtained by the second integral of \eqref{sumform1AP}. Its obtainment  is more delicate and requires  many intermediate steps. Our final result for it is:
\begin{eqnarray}
	\label{W_function_in}
	W_b(x,x')&=&-\frac{2q}{(2\pi)^D}\sum_{n=-\infty}^{+\infty}e^{iqn(\varphi-\varphi')}\int d^{D-2}{k} \ e^{i{\vec{k}}\cdot( {\vec{r}}-{\vec{r}}')_{\parallel}} \int_{\sqrt{{\vec{k}}^2+m^2}}^\infty dz z \frac{{\bar{K}}_{q|n+\alpha|}(az)}{{\bar{I}}_{q|n+\alpha|}(az)}\nonumber\\
	&\times& \frac{I_{q|n+\alpha|}(zr)I_{q|n+\alpha|}(zr')}{\sqrt{z^2-{\vec{k}}^2-m^2}}}\cosh[{\sqrt{z^2-{\vec{k}}^2-m^2}(t-t') ] \ .
\end{eqnarray} 
As we can see, in the limit $a\rightarrow \infty $ for fixed $r,r^{\prime }$, $	W_b(x,x')$ vanishes and, hence, it remains only the term $W_{cs}(x,x')$, that corresponds to the Wightman function in the the geometry of a cosmic string without cylindrical boundary.

\subsection{Wightman function in the exterior region}
\label{ext_reg}
In the region outside the cylindrical shell, the radial function of the wave-function is given in terms of a combination of Bessel and Neumann functions \cite{Abra}, as
\begin{eqnarray}
	\label{W-function}
	W_{\nu_n}(\gamma r)=C_1J_{q|n+\alpha|}(\gamma r)+C_2Y_{q|n+\alpha|}(\gamma r) \  .
\end{eqnarray}

Due to the Robin boundary condition \eqref{Dirbc} on the shell, we obtain the relation
\begin{eqnarray}
C_2/C_1=-\bar{J}_{q|n+\alpha|}(\gamma a)/\bar{Y}_{q|n+\alpha|}(\gamma a) \  , 	
\end{eqnarray}
for the coefficients in \eqref{W-function}. 

So, we can write the solution as,
\begin{eqnarray}
	\label{Out_sol}
\phi_\sigma(x)=\beta_\sigma g_{\nu_n}(\gamma r,\gamma a)e^{-i(\omega t-iqn\phi -i{\vec{k}}\cdot {\vec{r}}_{\parallel })} \ ,
\end{eqnarray}
where we have introduced the function
\begin{equation}
	\label{g-function}
	g_{q|n+\alpha|}(u,v)=J_{q|n+\alpha|}(v)\bar{Y}_{q|n+\alpha|}(u)-\bar{J}_{q|n+\alpha|}(u)Y_{q|n+\alpha|}(v) \ .
\end{equation}
Due to the  continuous values assumed by the quantum number $\gamma$, the normalization condition \eqref{Norm1}, provides
\begin{equation}
	|\beta_\sigma|^2=\frac{q\gamma}{2\omega (2\pi)^{D-1}}\frac 1{\bar{J}_{q|n+\alpha|}^2(\gamma a)+\bar{Y}_{q|n+\alpha|}^2(\gamma a)} \ .
	\label{beta2}
\end{equation}

Substituting \eqref{Out_sol} into the mode-sum formula \eqref{W.function}, the positive frequency Whightman function can be expressed as, 
\begin{eqnarray}
W(x,x')=\sum_\sigma |\beta_\sigma|^2 	g_{q|n+\alpha|}(\gamma r,\gamma a)	g_{q|n+\alpha|}(\gamma r',\gamma a)e^{-i[\omega(t-t')-qn(\varphi-\varphi')-{\vec{k}}\cdot( {\vec{r}}-{\vec{r}}')_{\parallel} ]}  \ ,
\end{eqnarray}
where, for this region, the summation is given by,
\begin{equation}
	\sum_{\sigma }=\sum_{n=-\infty}^{+\infty} \ \int d^{D-2}{k} \ \int d\gamma\ .  \label{Sumsig_1}
\end{equation}

Finally, taking the explicit expression for $|\beta_\sigma|^2$,  we can find the identity below:
\begin{equation}
\frac{g_{\nu_n}(\gamma r,\gamma a)g_{\nu_n}(\gamma r^{\prime },\gamma a)}{\bar{J}_{\nu_n}^{2}(\gamma a)+\bar{Y}_{\nu_n}^{2}(\gamma a)}=J_{\nu_n}(\gamma r)J_{\nu_n}(\gamma r^{\prime })-\frac{1}{2}\sum_{l=1}^{2}\frac{\bar{J} _{\nu_n}(\gamma a)}{{\bar H}_{\nu_n}^{(l)}(\gamma a)}H_{\nu_n}^{(l)}(\gamma r)H_{\nu_n}^{(l)}(\gamma r^{\prime }),  \label{relext}
\end{equation}%
where $H_{\nu}^{(l)}(z)$, $l=1,2$ are the Hankel functions \cite{Abra}. This allows us to present the Wightman function in the form \eqref{W_csb}, with its boundary-induced part being given by,
\begin{eqnarray}
W_b(x,x')&=&-\frac{q}{4(2\pi)^{D-1}}\sum_{l=1}^{2}\sum_{n=-\infty}^{+\infty} e^{iqn(\varphi-\varphi')}\int d^{D-2}{k} \ e^{i{\vec{k}}\cdot( {\vec{r}}-{\vec{r}})_{\parallel} }
 \int_0^\infty d\gamma \frac\gamma{\sqrt{\gamma^2+{\vec{k}}^2+m^2}}\nonumber\\
 &\times&\frac{\bar{J} _{\nu_n}(\gamma a)}{{\bar H}_{\nu_n}^{(l)}(\gamma a)}H_{\nu_n}^{(l)}(\gamma r)H_{\nu_n}^{(l)}(\gamma r^{\prime })e^{-i\sqrt{\gamma^2+{\vec{k}}^2+m^2}(t-t') } \  , \ \nu_n=q|n+\alpha| \ .
\end{eqnarray}
Now we can proceed in our development, by rotating the integral contour on the complex plane $\gamma $, as following:  we rotate the integration contour by the angle $\pi /2$ for $l=1$ and by the angle $-\pi /2$ for $l=2$. At this point we use the relations involving Bessel functions with imaginary arguments \cite{Abra}, finally  we divide the integrals over $\gamma$ in two segments: $[0, \ \sqrt{{\vec{k}}^2+m^2}]$ and  $[\sqrt{{\vec{k}}^2+m^2}, \ \infty )$. The integrals over the first segment cancel each other, remaining only the integrals over the second one. So, after some minors steps, we obtain for the boundary-induced Wightamn function the following expression:
 \begin{eqnarray}
\label{W_function_out}
W_b(x,x')&=&-\frac{2q}{(2\pi)^D}\sum_{n=-\infty}^{+\infty}e^{iqn(\varphi-\varphi')}\int d^{D-2}{k} \ e^{i{\vec{k}}\cdot( {\vec{r}}-{\vec{r}}')_{\parallel}} \int_{\sqrt{{\vec{k}}^2+m^2}}^\infty dz z \frac{{\bar{I}}_{q|n+\alpha|}(az)}{{\bar{K}}_{q|n+\alpha|}(az)}\nonumber\\
&\times& \frac{K_{q|n+\alpha|}(zr)K_{q|n+\alpha|}(zr')}{\sqrt{z^2-{\vec{k}}^2-m^2}}}\cosh[{\sqrt{z^2-{\vec{k}}^2-m^2}(t-t') ] \ .
 \end{eqnarray} 
As we can observe, the above boundary-induced Wightman function is obtained from the corresponding function for the interior region, \eqref{W_function_in},  by the replacements $I\rightleftarrows K$. Moreover, in the limit $a\rightarrow 0 $ for fixed $r,r^{\prime }$, \eqref{W_function_out} vanishes.

\section{Boundary-induced current}
\label{sec3}
The general expression for the bosonic current density operator is given by
\begin{eqnarray}
	\hat{j}_{\mu }(x)&=&ie\left[{\hat\phi} ^{\dagger}(x)D_{\mu }\hat{\phi} (x)-
	(D_{\mu }\hat{\phi)}^{\dagger}\hat{\phi}(x)\right] \nonumber\\
	&=&ie\left[\hat{\phi}^{\dagger}(x)\partial_{\mu }\hat{\phi} (x)-{\hat\phi}(x)
	(\partial_{\mu }\hat{\phi(x))}^{\dagger}\right]-2e^2A_\mu(x)|\hat{\phi}(x)|^2 \   .
	\label{J.mu}
\end{eqnarray}

Because, in this section, we are mainly interested to calculate the vacuum expectation value of boundary-induced current for regions inside and outside the cylindrical shell, we can evaluate them in terms of the corresponding positive frequency boundary-induced Wightman function as shown below:
\begin{equation}
	\label{Induced_current}
	\langle {\hat{j}}_{\mu}(x) \rangle_b=ie\lim_{x'\rightarrow x}
	\left\{(\partial_{\mu}-\partial_{\mu '})W(x,x')_b+2ieA_\mu W(x,x')_b\right\} \ .
\end{equation}

For the system under consideration, the only nonzero component of the current density is the azimuthal one. In this way,
we will focus only on the evaluation of this component. 

Writing, $	A_\varphi=-\frac{q\Phi}{2\pi}=q\frac{\alpha}{e} $ the boundary-induced azimuthal current reads,
\begin{eqnarray}
\label{b-Curr}
\langle {\hat{j}}_{\varphi}(x) \rangle_b=2ie\lim_{x'\rightarrow x}	\left\{\partial_{\varphi}W(x,x')_b+iq\alpha W(x,x')_b\right\} \ .
\end{eqnarray}

\subsection{Boundary-induced current inside the shell}
The boundary induced current in the region inside the shell is given by substituting \eqref{W_function_in} into \eqref{b-Curr}, after some intermediate steps, we obtain,
\begin{eqnarray}
	\label{Curr_in}
\langle {\hat{j}}_{\varphi}(x) \rangle_{b}^{(in)}&=&\frac{4qe}{(2\pi)^D}\sum_{n=-\infty}^{+\infty}q(n+\alpha)\int d^{D-2}{k} \nonumber\\
&\times& \int_{\sqrt{{\vec{k}}^2+m^2}}^\infty dz z \frac{{\bar{K}}_{q|n+\alpha|}(az)}{{\bar{I}}_{q|n+\alpha|}(az)}\frac{I^2_{q|n+\alpha|}(zr)}{\sqrt{z^2-{\vec{k}}^2-m^2}} \ . 
\end{eqnarray}
The integral over ${\vec{k}}$ can be developed with the help of the identity below \cite{Mello}, 
\begin{eqnarray}
	\int d^{D-2}\mathbf{k}\int_{\sqrt{k^{2}+m^{2}}}^{\infty }dz\frac{k^{s}f(z)}{\sqrt{z^{2}-k^{2}-m^{2}}}&=&\frac{\pi ^{\frac{D-2}{2}}}{\Gamma (\frac{D-2}{2})}B\left( \frac{D-2+s}{2	},\frac{1}{2}\right)\nonumber\\
	&\times& \int_{m}^{\infty }dz\,\left( z^{2}-m^{2}\right) ^{\frac{D-2+s-1}{2}}f(z),  \label{intk}
\end{eqnarray}
where $B(x,y)$ is the Euler beta function \cite{Grad}. Using this identity, we get:
\begin{eqnarray}
\label{Curr_in_1}	
\langle {\hat{j}}_{\varphi}(x) \rangle_{b}^{(in)}& =&\frac{8q^2ea^{1-D}}{(4\pi)^{\frac{D+1}{2}}\Gamma(\frac{D-1}{2})}\sum_{n=-\infty}^{+\infty}(n+\alpha)\int_{am}^\infty dv v (v^2-a^2m^2)^{\frac{D-3}{2}}\nonumber\\
&\times&\frac{{\bar{K}}_{q|n+\alpha|}(v)}{{\bar{I}}_{q|n+\alpha|}(v)}I^2_{q|n+\alpha|}((r/a)v) \  ,
\end{eqnarray}
where we have introduced a new integral variable by $v=az$. Considering the asymptotic expressions for the modified Bessel functions for large arguments \cite{Abra}: $I_\nu(z)\approx e^{z}/{\sqrt{2\pi z}}$, and $K_\nu(z)\approx \sqrt{\pi/(2z)}e^{-z}$, we can verify that for $r<a$ the integral in above expression is exponentially convergent in the upper limit, and the current is finite. 

Also it is possible to see that this current is an odd function of $\alpha$. We can obtain a more convenient expressions for \eqref{Curr_in_1} and \eqref{Curr_out} by writing the parameter $\alpha$  defined in \eqref{def1} in the form
\begin{equation}
	\alpha=n_0 +\alpha_0 \ \ {\rm with  \ |\alpha_0| \ < 1/2},
	\label{alphazero}
\end{equation}
where $n_0$ is an integer number. So redefining the quantum number $n$ in order to absorb $n_0$, both expressions will depend only on the fractional part, $\alpha_0$, of the ratio between the total magnetic flux, $\Phi$, by the quantum one, $\frac{2\pi}{e}$.   

Near the cylindrical shell \eqref{Curr_in_1} diverges, and this behavior is given for large values of $n$. In this limit, we will take, in the order of modified Bessel function, our first approximation: $n+\alpha_0\approx n$. Due to the presence of $\alpha_0\neq 0$, it remains a non-vanishing contribution in the summation over $n$, proportional to this parameter. Accepting this first order approximation, the procedure adopted to obtain a estimative of this divergent behavior is similar what was done in \cite{Mello}.  It is introduced a new integration variable $z\rightarrow nqz$, replacing the modified Bessel functions by their uniform asymptotic expansions for large values of the order \cite{Abra}, and expanding over $a-r$, up to the leading order, one finds
\begin{eqnarray}
	\label{Currnt_in_div}
\langle {\hat{j}}_{\varphi}(x) \rangle_{b}^{(in)} &\approx& \frac{q^De\alpha_0(2\delta_{B0}-1)}{2^{D-2}a^{D-1}\pi ^{\frac{D+1}{2}}\Gamma \left( \frac{D-1}{2}\right) }\int_{0}^{\infty }dz\frac{z^{D-2}}{\sqrt{1+z^{2}}}\sum_{n=1}^{\infty
	}n^{D-2}e^{-2nq(1-r/a)\sqrt{1+z^{2}}}  \  .
\end{eqnarray}
After developing, approximately, the sum over $n$ and performing the integral in the variable $z$,  we arrive to: 
\begin{eqnarray}
		\label{Currnt_in_div_1}
\langle {\hat{j}}_{\varphi}(x) \rangle_{b}^{(in)}\approx\frac{qe\alpha_0(2\delta_{B0}-1)}{2^D\pi^{\frac{D+1}{2}}}\frac{\Gamma \left( \frac{D-1}{2}\right)}{(a-r)^{D-1}} \  .
\end{eqnarray}

\begin{figure}[!htb]
	\begin{center}
		\includegraphics[scale=0.25]{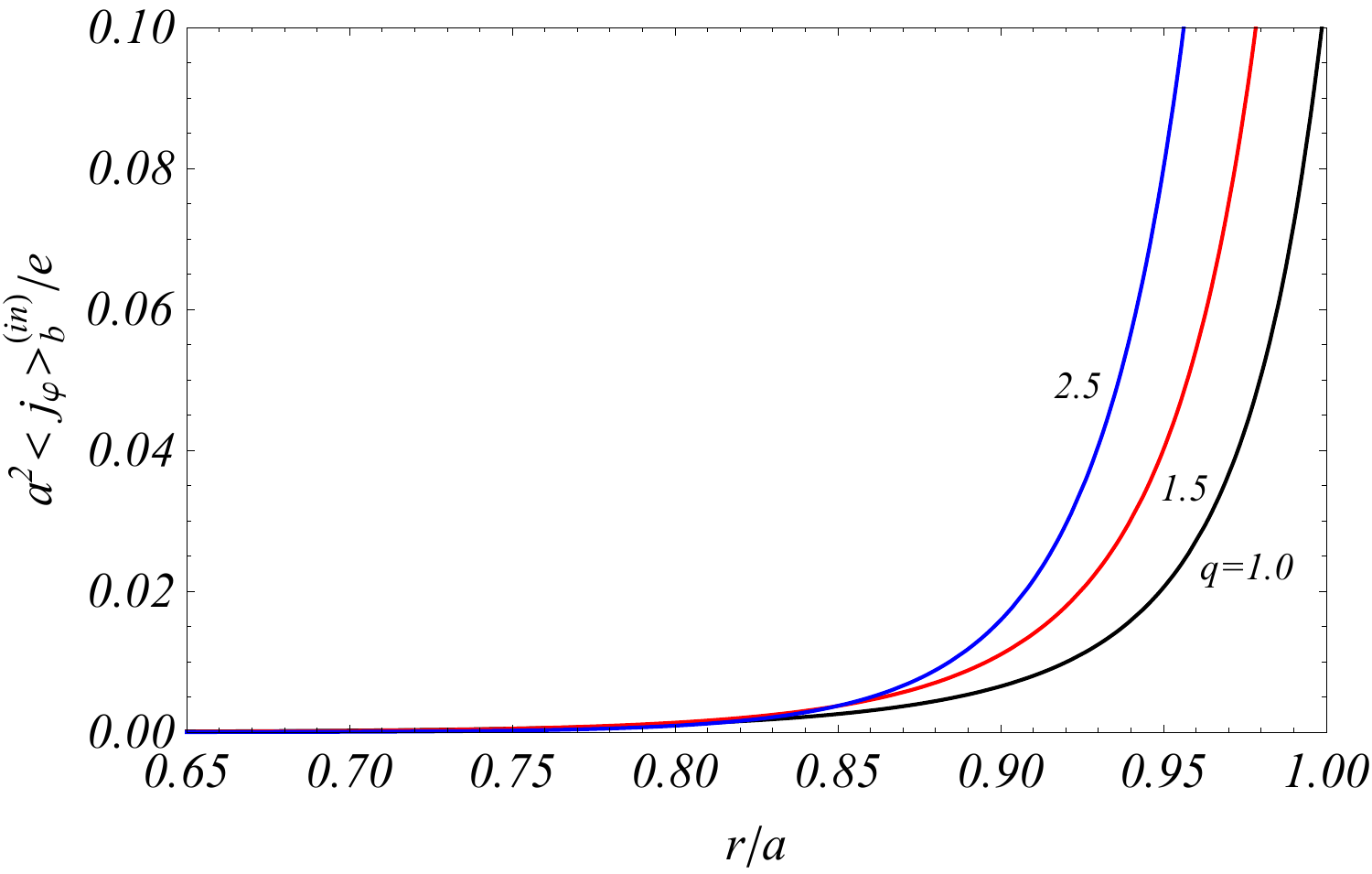}
		\quad
		\includegraphics[scale=0.35]{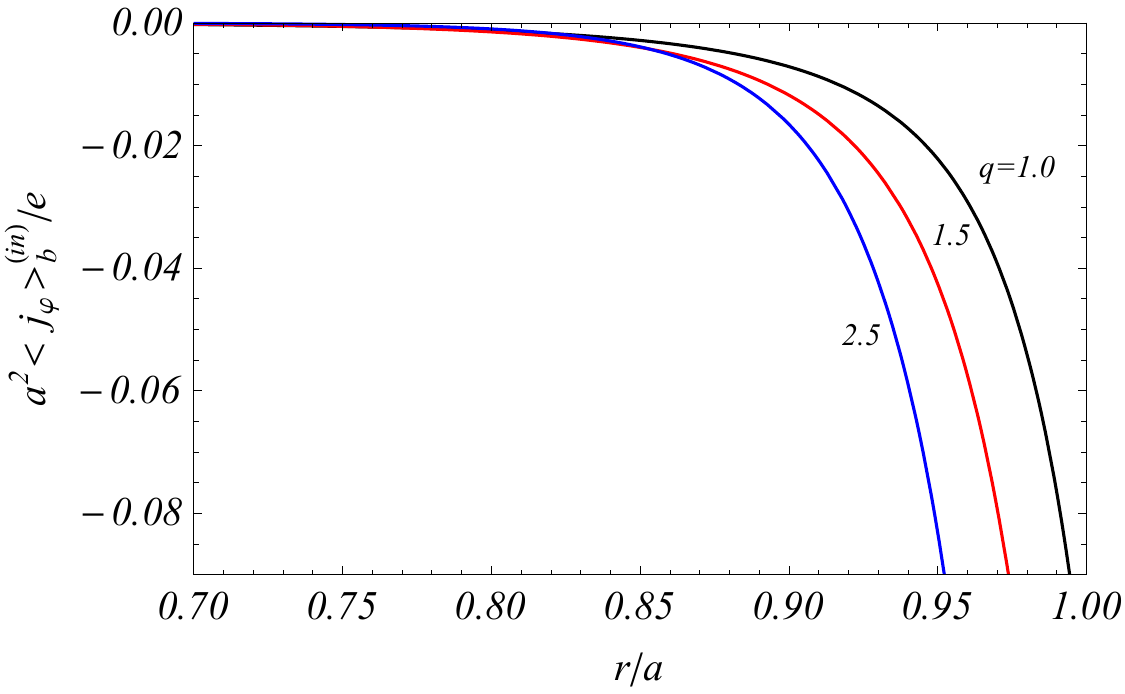}
		\caption{These plots present the behavior of the boundary induced currents densities for the region inside the shell in units of $e/a^2$, as function of $r/a$, considering Dirichlet boundary condition (left plot) and Neumann boundary condition (right plot). Different values of $q$ (the numbers near the curves) are adopted. Both graphs are plotted for $D=3$,  $\alpha_0=1/4$ and $ma=4$.}
		\label{fig1}
	\end{center}
\end{figure}

\begin{figure}[!htb]
	\begin{center}
		\includegraphics[scale=0.11]{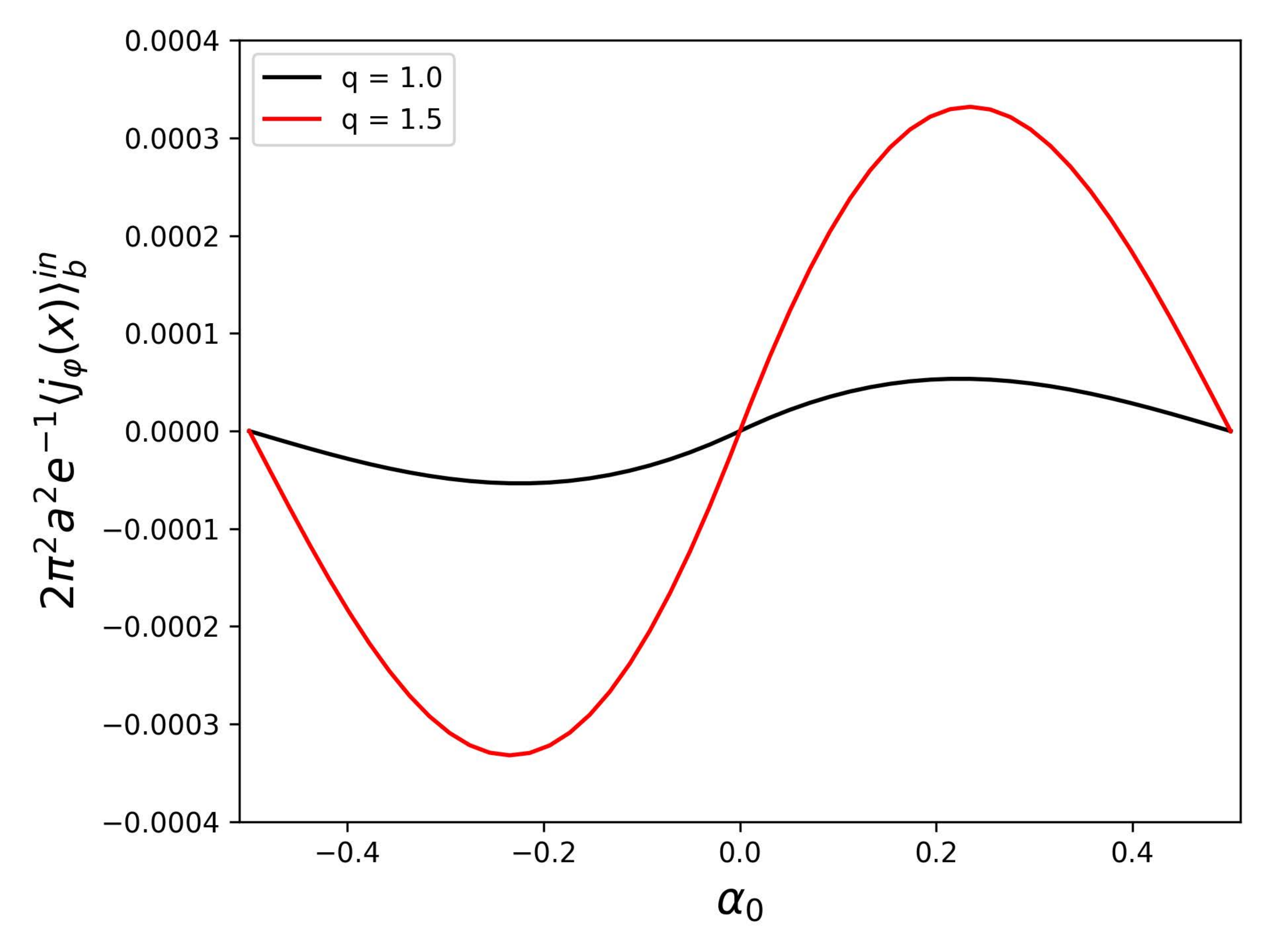}
		\caption{This plot exhibits the boundary induced currents densities in units of $e/a^2$ multiplied by $2\pi^2$, in the region inside the shell as function of $\alpha_0$ considering only Dirichlet boundary condition. We have adopted $D=3$,  $ma=4$, $r/a=0.5$ and two different values of $q$.}
		\label{fig1a}
	\end{center}
\end{figure}

In Fig. \ref{fig1} are exhibited the behaviors of the boundary induced currents in the region inside the shell as function of $r/a$, considering $D=3$, $\alpha_0=1/4$ and $ma=4$, for different values of the parameter associated with the planar angle deficit, $q$. In the left plot we adopted the Dirichlet boundary condition, and in the right the Neumann boundary condition. Moreover, in Fig. \ref{fig1a} we present the behavior of $\langle {\hat{j}}_{\varphi}(x) \rangle_{b}^{(in)}$ as function of $\alpha_0$ for $D=3$, considering $ma=4$ and $r/a=0.5$, for $q=1$ and $q=1.5$. In this plot we adopted just Dirichlet boundary condition. 

\subsection{Boundary-induced current outside the shell}
The boundary-induced current in the region outside the shell is obtained straightforward by the change  $I\rightleftarrows K$. The final result is,
\begin{eqnarray}
	\label{Curr_out}	
	\langle {\hat{j}}_{\varphi}(x) \rangle_{b}^{(out)} &=&\frac{8q^2ea^{1-D}}{(4\pi)^{\frac{D+1}{2}}\Gamma(\frac{D-1}{2})}\sum_{n=-\infty}^{+\infty}(n+\alpha)\int_{am}^\infty dv v (v^2-a^2m^2)^{\frac{D-3}{2}}\nonumber\\
	&\times&\frac{{\bar{I}}_{q|n+\alpha|}(v)}{{\bar{K}}_{q|n+\alpha|}(v)}K^2_{q|n+\alpha|}((r/a)v) \  .
\end{eqnarray}
Here, we can also verify, by using the asymptotic expansions of the modified Bessel functions,  that for points for $r>a$ this current is finite.

As to the region inside the boundary, \eqref{Curr_out} diverges near the cylindrical shell. The leading term can be obtained in the same way as in the last analysis. The corresponding asymptotic expansion is similar with \eqref{Currnt_in_div}, by changing $(a-r)$ by $(r-a)$. Now we will analyze the behavior of \eqref{Curr_out} for large distances from the cylindrical surface, $r\gg a$. Let us start first with the massless scalar field case first.\footnote{In this analysis we assume that we have  adopted the notation \eqref{alphazero}, and redefined the summation over $n$.} In order to develop this analysis we introduce a new integration variable $y=zv/a^2$ and expand the integrand in powers of $a/r$. Because we are assuming $|\alpha_0|<1/2$, the main contribution comes from $n=0$. Doing this we have to integrate the square of  MacDonald function with the help of \cite{Grad}. After some intermediate steps, we obtian:  
\begin{eqnarray}
\label{asymptotic_1}
\langle {\hat{j}}_{\varphi}(x) \rangle_{b}^{(out)}\approx \frac{|\alpha_0|}{\alpha_0}\frac{C_D}{a^{D-1}}\left(\frac{2a}{r}\right)^{D-1+2q|\alpha_0|} \  , 
\end{eqnarray}
where
\begin{eqnarray}
C_D=\frac{2qe}{4^{q|\alpha_0|}(4\pi)^{\frac{D+1}{2}}}\frac{(B|\alpha_0|q+Aa)}{(-B|\alpha_0|q+Aa)}\frac{\Gamma\left(\frac{D-1}{2}+2q|\alpha_0|\right)\left(\Gamma\left(\frac{D-1}{2}+q|\alpha_0|\right)\right)^2}{\Gamma\left(q|\alpha_0|\right)^2\Gamma(D-1+2q|\alpha_0|)} \ .
\end{eqnarray}

For the case of massive field, such that $mr\gg 1$, and considering also that $am\gg1$, the main contribution into the integral in (\ref{Curr_out}) is given  from the lower limit. As to the square of the Macdonald function we take its asymptotic form \cite{Abra}. So after some minor intermediate steps, we obtain:
\begin{equation}
	\label{asymptotic_2}
\langle {\hat{j}}_{\varphi}(x) \rangle_{b}^{(out)}\approx {\frac {2e\pi{q}^{2}{m}^{\frac{D-3}{2}}{{\rm e}^{-2\,mr}}}{ \left( 4\pi\right)^{\frac{D+1}2}{r}^{\frac{D+1}2}}}\sum_{n=-\infty}^{+\infty}(n+\alpha_0)\frac{\bar{I}_{q|n+\alpha_0|}(ma)}{\bar{K}_{q|n+\alpha_0|}(ma)} \  ,
\end{equation}
where there is an exponential decay.

\begin{figure}[!htb]
	\begin{center}
		\includegraphics[scale=0.27]{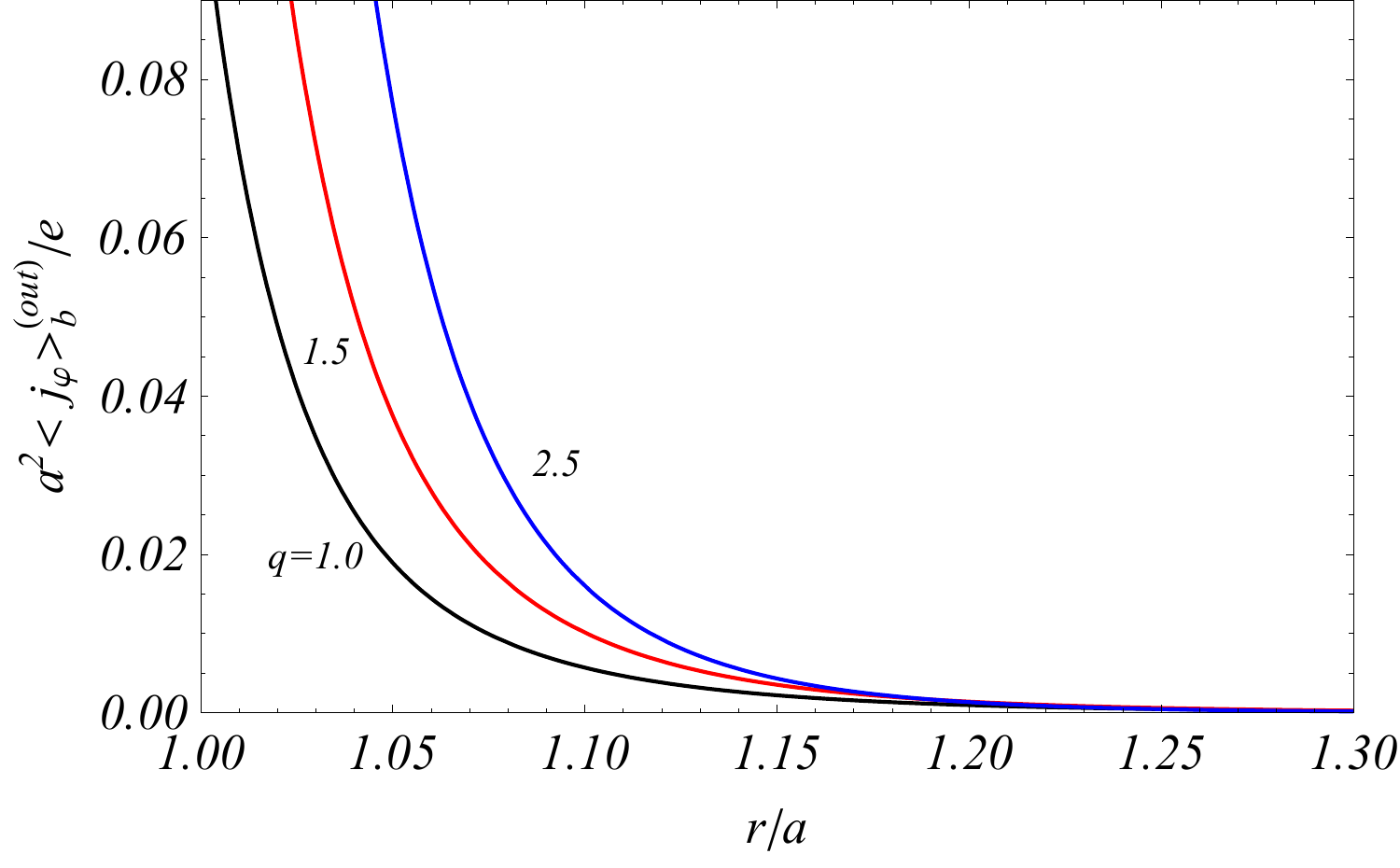}
		\quad
		\includegraphics[scale=0.36]{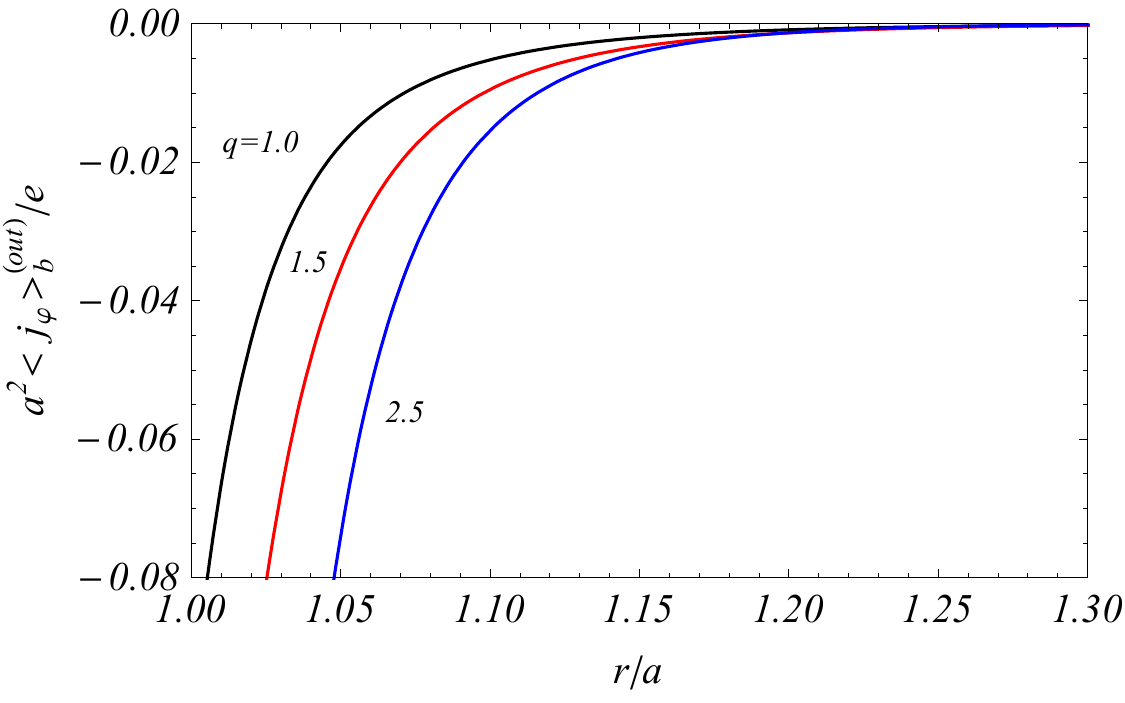}
		\caption{These plots present the behavior of the boundary induced currents densities in units of $e/a^2$, as function of $r/a$ for the region outside the shell and considering Dirichlet boundary condition (left plot) and Neumann boundary condition (right plot). Different values of $q$ (the numbers near the curves) are adopted. Both graphs are plotted for $D=3$,  $\alpha_0=1/4$ and $ma=4$.}
		\label{fig2}
	\end{center}
\end{figure}

\begin{figure}[!htb]
	\begin{center}
		\includegraphics[scale=0.25]{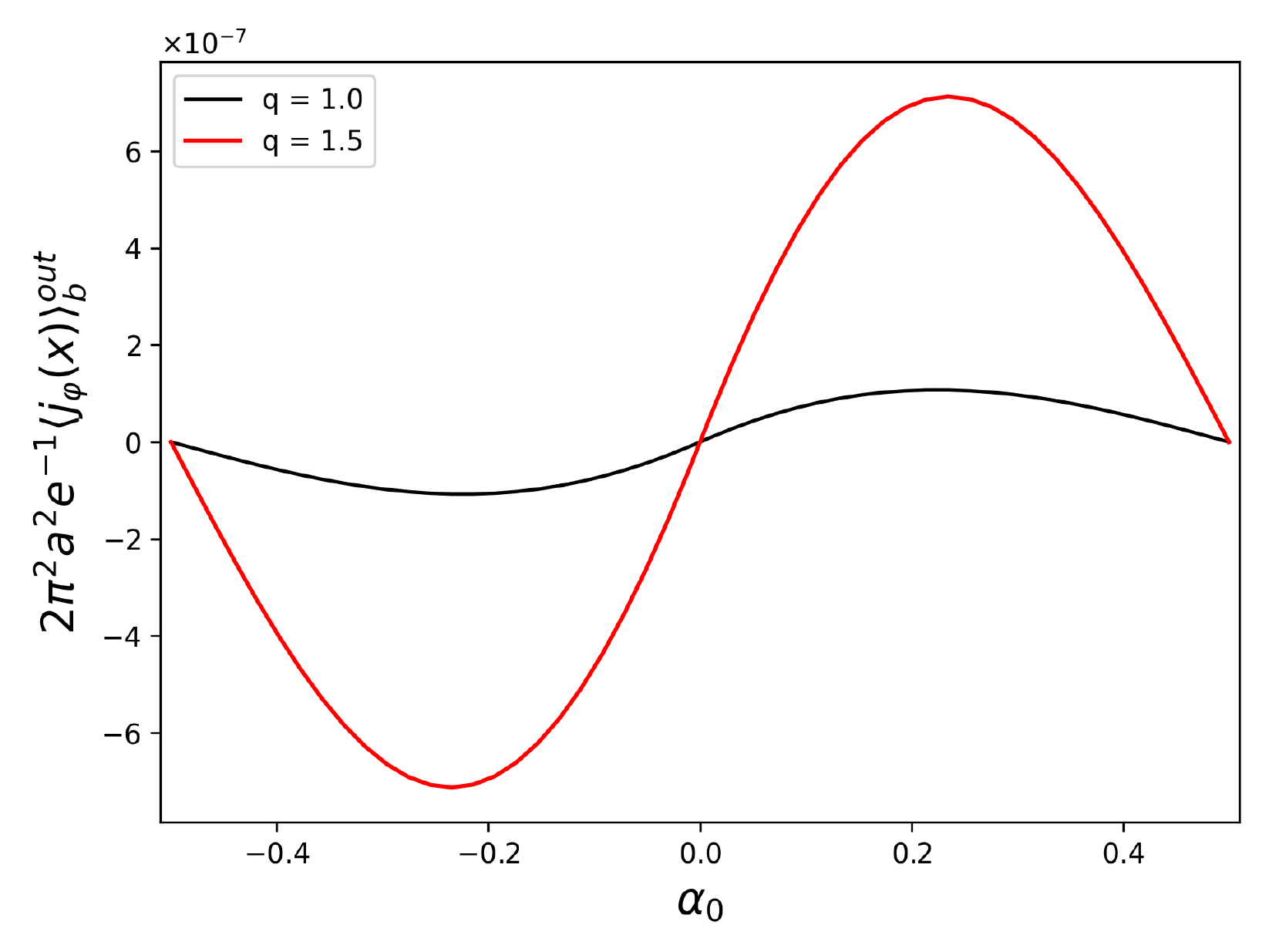}
		\caption{This plot exhibits the behavior of the boundary induced currents densities in units of $e/a^2$ multiplied by $2\pi^2$ in the region outside the shell as function of $\alpha_0$ considering only Dirichlet boundary condition. Here we have adopted $D=3$,  $ma=4$, $r/a=1.5$ and two different values of $q$.}
		\label{fig2a}
	\end{center}
\end{figure}

In Fig. \ref{fig2} we present the behaviors of the boundary induced currents in the region outside the shell as function of $r/a$, considering $D=3$, $\alpha_0=1/4$ and $ma=4$, for different values of the parameter associated with the planar angle deficit, $q$. In the left plot we assumed Dirichlet boundary condition, and in the right Neumann boundary condition.  Also, in Fig. \ref{fig2a} we exhibit the behavior of $\langle {\hat{j}}_{\varphi}(x) \rangle_{b}^{(out)}$ as function of $\alpha_0$ for $D=3$, considering $ma=4$ and $r/a=1.5$, for $q=1$ and $q=1.5$. Here only Dirichlet boundary condition was considered.  

\section{Conclusions}
\label{conc} 
In this work, we have analyzed the vacuum bosonic current  in a general $(1+D)-$dimensional cosmic string spacetime induced by the magnetic flux running along its core and taking into account the presence of a cylindrical boundary, of radius $a$,  coaxial to it. We assumed that the quantum field obeys the Robin boundary condition on the cylindrical surface. An important mathematical quantity used to calculate the induced current is the Wightman function. This function is expressed in terms of a sum over the complete set of normalized solutions of the Klein-Gordon equation, as given in \eqref{W.function}. In order to obtain this function, we have calculated bosonic wave-function compatible with the boundary condition, Eq. \eqref{Dirbc}, for both regions of the space. In our development we expressed the Wightman functions, as the sum of a boundary-free term plus a boundary-induced one. Because the analysis of the boundary-free vacuum bosonic current has been exhaustively analyzed in literature (see \cite{LS}, \cite{SNDV} and \cite{Braganca2015}), our main objective here is to investigate the boundary-induced current.  With this objective, in sections \ref{int_reg} and \ref{ext_reg} we presented the boundary induced Wighman functions, for the regions inside and outside the shell, respectively. The calculations of the boundary-induced currents were developed in section \ref{sec3}. There, we mention that for the system under consideration, only azimuthal currents densities are induced. 

For the region inside the shell, we obtained the expression \eqref{Curr_in_1} in an integral representation. We shown that for points near the string's core, $r<a$, the integral is exponentially convergent in the upper limit, and the current is finite. On the other hand near the shell, $r\leq a$, the current diverges with $\frac1{(a-r)^{D-1}}$, as exhibited in \eqref{Currnt_in_div}. In order to provide a better understanding of the behavior of this current, in Fig. \ref{fig1}, we presented its plots as function of $r/a$, adopting $D=3$,  $\alpha_0=1/4$ and $ma=4$, considering, separately, the Dirichlet and Neumann boundary conditions. Different values of the parameter associated with the planar angle deficit, $q$, were used. As we can see the intensity of the boundary induced current increases  with  $q$; moreover, it presents different signs when Dirichlet or Neumann conditions are considered. In addition, in Fig. \ref{fig1a}, we exhibit the behavior of the of $\langle {\hat{j}}_{\varphi}(x) \rangle_{b}^{(in)}$ as function of $\alpha_0$ considering only Dirichlet boundary condition.

For the region outside the shell the boundary-induced current was given in \eqref{Curr_out}. We have also shown that for points near the boundary the azimuthal current diverges as $\frac{1}{(r-a)^{D-1}}$.  The behavior of the current for points with $r>>a$, was analyzed for  massless and massive fields. For massless field the current presents a power-like decay given by \eqref{asymptotic_1}; as to massive case the current decay exponentially, $e^{-2mr}$, as exhibited in  \eqref{asymptotic_2}. We also provide for this region the plots exhibiting the behavior of this current with $r/a$ in Fig. \ref{fig2}. In these plots we considered $D=3$,  $\alpha_0=1/4$ and $ma=4$, admitting that field obeys Dirichlet and Neumann boundary conditions on the cylindrical shell. For this case we observe that the current presents the same characteristics as mentioned before, about its behavior with $q$ and boundary conditions. The Fig. \ref{fig2a}, presents the behavior of the of $\langle {\hat{j}}_{\varphi}(x) \rangle_{b}^{(out)}$ as function of $\alpha_0$ considering only Dirichlet boundary condition.

Also we would like to mention that the boundary-free azimuthal current diverges near the idealized string's core as $\frac1{r^{D-1}}$ (see \cite{Braganca2015}). So, in this region the total induced current is dominated by the boundary-free part. On the other hand, for points near the shell, the boundary-free bosonic current is finite, and the total currents are dominated by the boundary-induced contritions. Moreover, for points far from the string, i.e., $r>>a$, and considering massless field, the boundary-free azimuthal current decays as power law, $\frac1{r^{D-1}}$, and the boundary-induced one as $\frac1{r^{D-1+2q|\alpha_0|}}$, as shown in \eqref{asymptotic_1}. For massive field, the boundary-free decays exponentially with $e^{-2mr\sin(\pi/q)}$ and boundary-induced ones decay as $e^{-2mr}$.

To finish this paper we would like to say that a possible application of the present calculations is in condensed matter physics. In \cite{Kleman}, Kleman has shown that  disordered solids or liquid crystals, may present linear topological defects, named dislocation or disclination.  In their seminal work, Katanaev and Volovich \cite{Katanaev} have shown that there is a strong geometrical similarity between disclination and cosmic string, and for some applications both kind of linear defects may be dealt with through the same geometric methods.\footnote{In \cite{Osipov} Osipov has shown that the change in the topology of a medium introduced by a linear defect as a disclination in an elastic solid or cosmic string in space-time has strong effects on the physical properties of the medium.}  In this sense the study of a possible azimuthal current in a liquid crystal confined in a cylindrical shell having a magnetic flux running along a disclination in the center of this container may be used to corroborate the results presented in this paper. 
 We also note that the change in sign of the induced current under variation of boundary conditions (Dirichlet vs Neumann) provides a clear signature of the sensitivity of vacuum polarization to boundary physics. This effect could, in principle, be observed in condensed matter analogues under controlled laboratory conditions.
\section*{Acknowledgment}
We would like to thank Aram A. Saharian for  helpful discussions.  H.F.S.M. is partially supported by the
Brazilian agency CNPq under Grant No. 308049/2023-3.

\end{document}